\documentclass[11pt,fleqn]{article}
\title{Comments on "Anomalous Hydrodynamic Drafting of \\ Interacting Flapping Flags"}
\author{Guanghua Zhu} 
\date{Qinghaihu Centre, 2, Street Hua-Yuan South,
 Xining 810000, China} 
\begin{document}           
\maketitle
\begin{abstract}
We make remarks on  Ristroph and Zhang's [{\it Phys. Rev. Lett.}
{\bf 101}, 194502  (2008)] paper. We argue especially that due to
the interferences the calibration procedures in [1] were not
complete and this will induce some measurements' error.
%
\end{abstract}
\doublerulesep=6.8mm    %
\baselineskip=6.8mm
\bibliographystyle{plain}
Ristroph and Zhang just presented an interesting observation about
the  inverted drag relationship for 6 tandem flags where inverted
drafting directly suggests hydrodynamic repulsion between flags
[1]. Unfortunately they only presented their observations about
the inverted drag relationship for 2 and 6 tandem flags
 [1]. They didn't illustrate the crucial results : 3
tandem flapping flags in [1]. There are inconsistencies as
evidenced from Fig. 2 and Fig. 4 (b), (c). In Fig. 2, $D/D_0 \sim
0.5, 1$ for the first flag while $D/D_0 \sim 1,1.5$ for the 2nd.
one considering the spacing $G/L=0,0.6$, respectively? In Fig. 4
(b), (c); $D/D_0 \sim 0.5, 0.8$ for the first flag while $D/D_0
\sim 0.7,1.3$ for the 2nd. one considering the spacing
$G/L=0,0.6$, respectively? Is the difference due to interferences
and upstream influences?\newline To experimentally capture the
forcing as well as model the interaction of shapechanging bodies,
they inserted thin flexible filaments into a flowing soap film
[2]. Each filament is fixed at its upstream end to a thin wire, a
flagpole that extends out of the film, while the rest of the
thread hangs free in the film. It means the conventional sting (a
{\it smart} support for force and moment measurements) in wind
and/or water tunnel has been moved from the downstream part of a
test section to the upstream part [3]. Although the diameter
(34$\mu$m? [2]) of the flagpole (or a sting) is small once it is
inserted transverse to the flow (in fact, a soap solution with air
interfaces along upper and lower fluctuating surfaces; film
thickness : 4.7 $\mu$m [1]) there will be wakes generated and
small vortices shedding. The later complex flow structure will
influence the downstream flow condition of one flexible flag (a
thread with diameter : 300 $\mu$m [1]) or tandem flags behind. To
be precise, the upstream flow is not essentially a uniform flow or
free of turbulence.
\newline To obtain the detailed drag-force characteristics it is
important to calibrate necessary sets of data beforehand. As
mentioned in [1], Ristroph and Zhang {\it only} measured the
time-averaged streamwise fluid force : $D_0$  or the drag on an
{\it isolated} flag [2] which is served as a normalization or a
baseline [1] for comparison. The present author argues that above
procedure is not enough for drawing a conclusion
about the six tandem flags. 
\newline
Normally the force or moment measurements conducted in wind or
water tunnels are for one-body or one-object (say, scaled airfoil
or aircraft model) [3]. Considering the multi-body purpose
(configuration) as well as the unsteady flow characteristics the
interferences and active or passive influences between different
bodies (located either forward or backward with respect to
(w.r.t.) a referenced body (say, $D_0$ in [1])) which are confined
in a limited test section should be carefully examined firstly. It
means, e.g., for 2 flags, the same measuring procedure should be
conducted for both configurations to eliminate the possible noises
: (a) to put flag 1  forward with flag 2  rearward, and (b) to put
flag 2  forward with flag 1  rearward. The measured drags are :
$D^f_{1f}$, $D^r_{1r}$, $D^r_{1f}$, and $D^f_{1r}$, respectively.
\newline To be specific, we can check whether $D^f_{1f}=D^r_{1f}$
or $D^r_{1r}=D^f_{1r}$?
For each flag of this case, Ristroph and Zhang thought that $D_0$
is the same for either the forward one or the rearward one.
However, as mentioned above, once there is a sting (flagpole) or a
flag mounted inside the flow, there are severe flow (or structure)
induced fluctuations (considering the force or moment induced)
generated. The second flag, no matter it will be located forward
or rearward to the already mounted one with a spacing in-between,
once being inserted into the flow, the final situation is
different from that two tandem flags being mounted simultaneously
into the flow! The present author proposes that we should compare
the above 4 drag forces for each flag  with $D_0$.  We believe
that the normalization or baseline drag force for two tandem flags
should be selected from either $D_{1f}$ or $D_{1r}$. The
subsequent results considering the same demonstrating procedure of
Fig. 2 in [1] should be different as the normalization drag is
different from $D_0$. With this process, we can minimize the
interferences and mutual influences for multi-body configurations.
Similar procedures must be conducted for $i\ge 3$ (w.r.t. $D_i$),
e.g., $i=5$ to check the validness of 6 tandem flags or Fig. 4
(b), (c) in [1].
\newline To conclude in brief, we believe the effects of interferences
and mutual influences could tune the conclusion made in [1]. The
preliminary direct check is presented above and the other check is
to put the next or following flag upstreamwise instead
downstreamwise when one flag has been inserted into the soap
solution.

\end{document}